\documentclass{nrc1}

\usepackage{graphicx}
\usepackage{amsmath}
\usepackage{amssymb}
\usepackage{bm}
\usepackage{subcaption}
\usepackage[english]{babel}

\title{A Meta-Analysis of LHC Results}

\author{Sevim A\c c\i{}ks\" oz}
\address[label1]{Bo\u{g}azi\c{c}i University, Department of Physics, Istanbul}
\author{Bilal \c Cark}
\address[label1]
\author{Selim Mert K\i{}rp\i{}c\i{}}
\address[label1]
\author{Merve Y\i{}ld\i{}z}
\address[label2]{Center for Axion and Precision Physics, IBS, Korea Advanced Institute of Science and Technology, Daejeon}
\author{Veysi Erkcan \"Ozcan}
\address[label1]

\IDnote{kahvelab@gmail.com}
\IDnote{Also at Feza G\"ursey Center for Physics \& Mathematics, Bo\u{g}azi\c{c}i University, Istanbul }

\shortauthor{A\c{c}{\i}ks\"oz, \c{C}ark, K\i{}rp\i{}c\i{}, Y\i{}ld\i{}z, \"Ozcan}

\begin{document}

\maketitle

\begin{abstract}
We report the preliminary results of a meta-analysis conducted to examine possible biases in the uncertainty values published in papers by ATLAS and CMS experiments. We have performed this analysis using two independent techniques; a vectoral analysis of the vector graphics files and a bitmap analysis of the raster graphic files of the exclusion plots from various physics searches. In both procedures, the aim is to compute the percentages of the data points scattered within 1-sigma and 2-sigma bands of the plots and verify whether the measured percentages agree with statistical norms assuming unbiased estimations of the uncertainties.

\end{abstract}

\section{Introduction}

The Large Hadron Collider experiments have published hundreds of articles and conference proceedings since they started collecting collision data about ten years ago. While many of these publications are on measurements of the Standard Model and the Higgs Boson, a large fraction relates to searches for new physics beyond the Standard Model (BSM), mainly supersymmetry and other \textit{exotic} models. None of those searches have yet yielded positive results, but provided constraints on the masses of hypothetical particles, their production cross-sections, and the coupling constants amongst them.

These constraints are often summarized as exclusion limits, obtained through hypothesis testing. The extraction of these limits are quite often a major portion of the analysis as they require a good deal of statistical modelling and a detailed understanding of the behaviour of the detectors. As such, the LHC collaborations form subgroups to set  policies regarding which statistical analysis libraries to use and regarding handling of a myriad of different systematic uncertainties from a multitude of sub-detectors.

But are these policies as sound as aimed? How successful are the
collaborations in properly estimating the systematic uncertainties? Are
the statistical analysis libraries indeed bug-free? Do the various
assumptions underlying the statistical techniques hold? For example, how
reliable are the ``asymptotic approximations'' used in many of these
analyses~\cite{cowan2011}? (See for instance the comparison of exclusion
limits derived from ensemble tests vs. asymptotic formulae in Figure~7
of~\cite{ATLASZgamma2017}.) In general, biased estimations can lead to
premature rejection of theoretical models or inflate the significance of
statistical fluctuations.

In this work, we present the preliminary results of a meta-analysis of 56 articles (Table~\ref{tab:articleList}), mostly on exotic BSM physics, from ATLAS and CMS Collaborations published between 2016 and 2018. Our analysis is conducted through two independent methods in order to minimize the possibility of introducing any unintentional biases ourselves. We re-digitize 141 ATLAS and 117 CMS exclusion plots, either in vector graphics form or in the form of bitmaps and then extract the percentages of the observed data points scattered within $1\sigma$ (green) and $2\sigma$ (yellow) uncertainty bands. We find that while the fraction of points falling into the $2\sigma$ band is consistent with 95\%, the fraction is higher than 68\% for the $1\sigma$ band, particularly for the CMS plots.

\section{Data}

Our dataset is composed of 95\% confidence-level exclusion plots, one example of which can be seen in Figure~\ref{fig:exclusion}. These plots show which parts of the parameter space is excluded by the data. Most often the horizontal axis represents an experimentally reconstructed variable, such as the mass of a theoretically predicted new particle, while the vertical axis is the cross section for the production of that particle. The observed limit is commonly plotted with a dark (black or dark blue) solid curve, and in the mass vs. cross-section plots it divides the parameter space into two regions with the above region excluded. In addition, plotted as a dashed curve, we see the expected limit, obtained from Monte Carlo pseudo-data produced with the assumption of no new physics.

The expected limit curve is surrounded by 1$\sigma$ (often green-colored) and $2\sigma$ (often yellow-colored) bands; indicating how far the expected limit can move due to the statistical fluctuations in the analyses. We expect to see about 68\% (95\%) of the  points that constitute the observed limit curve to be within the 1$\sigma$ ($2\sigma$) band. It is worth noting that quite often the curves are not smooth, with visible kinks at a number of locations. The reason for this is the way the analyses are conducted: during the production of Monte Carlo signal samples, a finite number of mass (or other model parameter) values are picked for the searched-for hypothetical particle, the cross-section limits are extracted for those discrete samples, and the final exclusion curves are constructed by extrapolation (so they are really piecewise linear functions). Our vector-graphics analysis focuses only on the data points at those kinks, while the bitmap analysis runs on all the pixels making up the exclusion curves.

\begin{table}[hbt!]
\centering
\topcaption{\label{tab:articleList}arXiv codes of the papers processed and
  the total number of data points that the vector graphics analysis could
  extract from the plots of each paper.}
\resizebox{1\textwidth}{!}{
\begin{tabular}{c|r|c|r|c|r||c|r|c|r}
\hline*    
\multicolumn{6}{c||}{CMS} & \multicolumn{4}{c}{ATLAS}\\\hline
1606.08076 &  244 & 1609.05391 & 70 & 1610.08066 & 58 & 1606.03833 & 1008 & 1606.03977 & 273\\
1611.06594 &   20 & 1612.05336 & 32 & 1612.09516 &894 & 1606.04833 &   42 & 1608.00890 &  78\\
1701.07409 &   40 & 1703.01651 & 45 & 1704.03366 & 30 & 1703.09127 &   37 & 1705.10751 &  13\\
1705.09171 &  239 & 1706.03794 &  7 & 1706.04260 & 60 & 1706.04786 &  351 & 1707.04147 & 598\\
1707.01303 &  108 & 1708.01062 & 20 & 1709.00384 & 26 & 1708.04445 &  174 & 1708.09638 & 470\\
1709.05406 &   99 & 1709.05822 & 18 & 1709.08908 & 68 & 1709.06783 &   18 & 1710.07235 & 220\\
1711.00431 &   10 & 1711.00752 & 14 & 1712.02345 &  8 & 1711.03301 &   48 & 1712.06518 & 141\\
1712.03143 & 2376 & 1801.03957 & 60 & 1802.01122 & 64 & 1801.07893 &   12 & 1801.08769 &  39\\
1802.02965 &   60 & 1803.03838 & 62 & 1803.06292 & 57 & 1802.03388 &  192 & 1803.09678 & 174\\
1803.10093 &  145 & 1803.11133 &340 & 1804.07321 & 27 & 1804.03496 &  304 & 1804.06174 &  66\\
1805.04758 &   44 & 1805.10228 &121 & 1806.00843 & 41 & 1804.10823 &  122 & 1805.01908 &2699\\
1806.05996 &   20 & & & & & & & &\\
\hline*    
\end{tabular}
}  
\end{table}

\section{Methods}

\subsection{Vector Graphics Analysis}

For the vector based analysis, EPS or PDF versions of the exclusion
plots are collected from the arXiv source files of the respective
articles, then converted to SVG file format using the open source program
\texttt{pdf2svg}~\cite{pdf2svg2008} in order to have a single
common XML interface to
uniformly parse them.  All the basic shapes in SVG files are described by
XML elements called `path's and various attributes of a path describe the
features of the object, such as its color, stroke-shape, etc.
Parsing using Python library svgpathtools~\cite{svgpathtools2018}, objects
with desired attributes can be filtered by element type as well as attributes
using standard regex pattern matching tools like \texttt{grep}.

The first filtering step is to
extract the paths representing $1\sigma$ and $2\sigma$ bands according to
their RGB decimal color codes.  The relevant RGB color codes are identified by
using the vector graphics editor program Inkscape~\cite{inkscape2015}.
During this inspection we have
discovered that while ATLAS seems to follow a standard color code for all
their plots (with very rare exceptions like the blue exclusion plots of
1804.03496, but ATLAS also provides the standard-colored versions as
auxiliary plots on their website~\cite{atlasexoticspublic}), CMS does
not have a single standard and thus the analyses
of their plots need special care (see Table~\ref{tab:colorcodes}).

\begin{table}[hbt!]
\centering
\topcaption{\label{tab:colorcodes}Different color codes used in ATLAS and CMS exclusion plots.}
\begin{tabular}{c|c|c}
\hline*    
Color & ATLAS & CMS \\\hline
Green &  rgb(0\%,100\%,0\%) & rgb(13.33313\%,54.508972\%,13.33313\%) \\
& & rgb(0\%,79.998779\%,0\%) \\
\hline
Yellow & rgb(100\%,100\%,0\%) & rgb(100\%,84.312439\%,0\%) \\
& & rgb(100\%,79.998779\%,0\%) \\
& & rgb(100\%,100\%,19.999695\%) \\
& & rgb(100\%,80.076599\%,0\%) \\
\hline*    
\end{tabular}
\end{table}

\par
\begin{figure*}[hbt!]
\centering
\setlength{\captionwidth}{0.\textwidth}
\topcaption{\label{fig:exclusion}Sample exclusion plot from 1606.03833, an ATLAS paper~\cite{atlas2016}, and the extracted `skeleton' of the needed lines.}
\begin{subfigure}[b]{0.49\textwidth}
\centering
\includegraphics[width=1.05\textwidth]{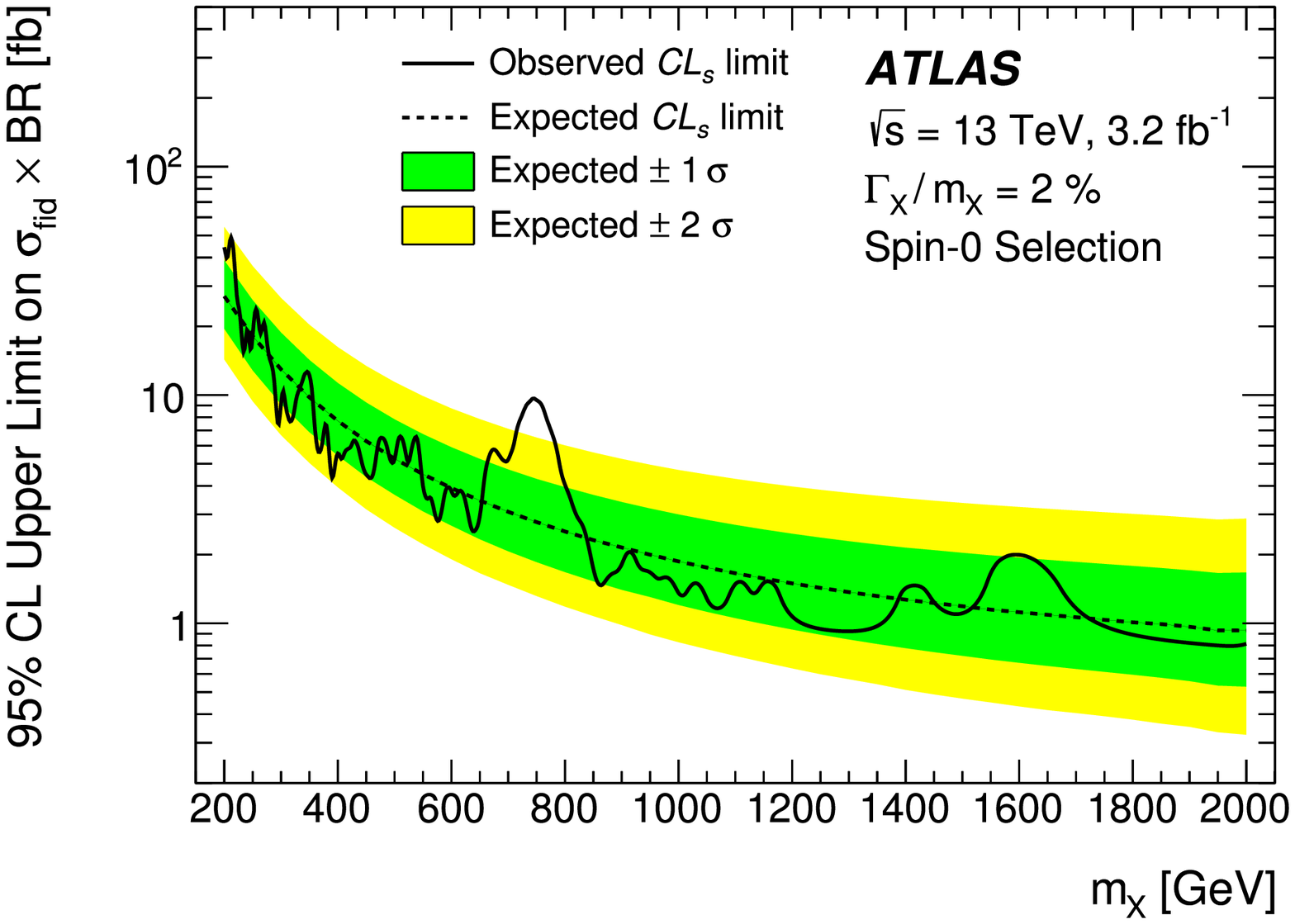}
 \caption{}
\end{subfigure}
\begin{subfigure}[b]{0.49\textwidth}
\centering
\includegraphics[width=0.88\textwidth]{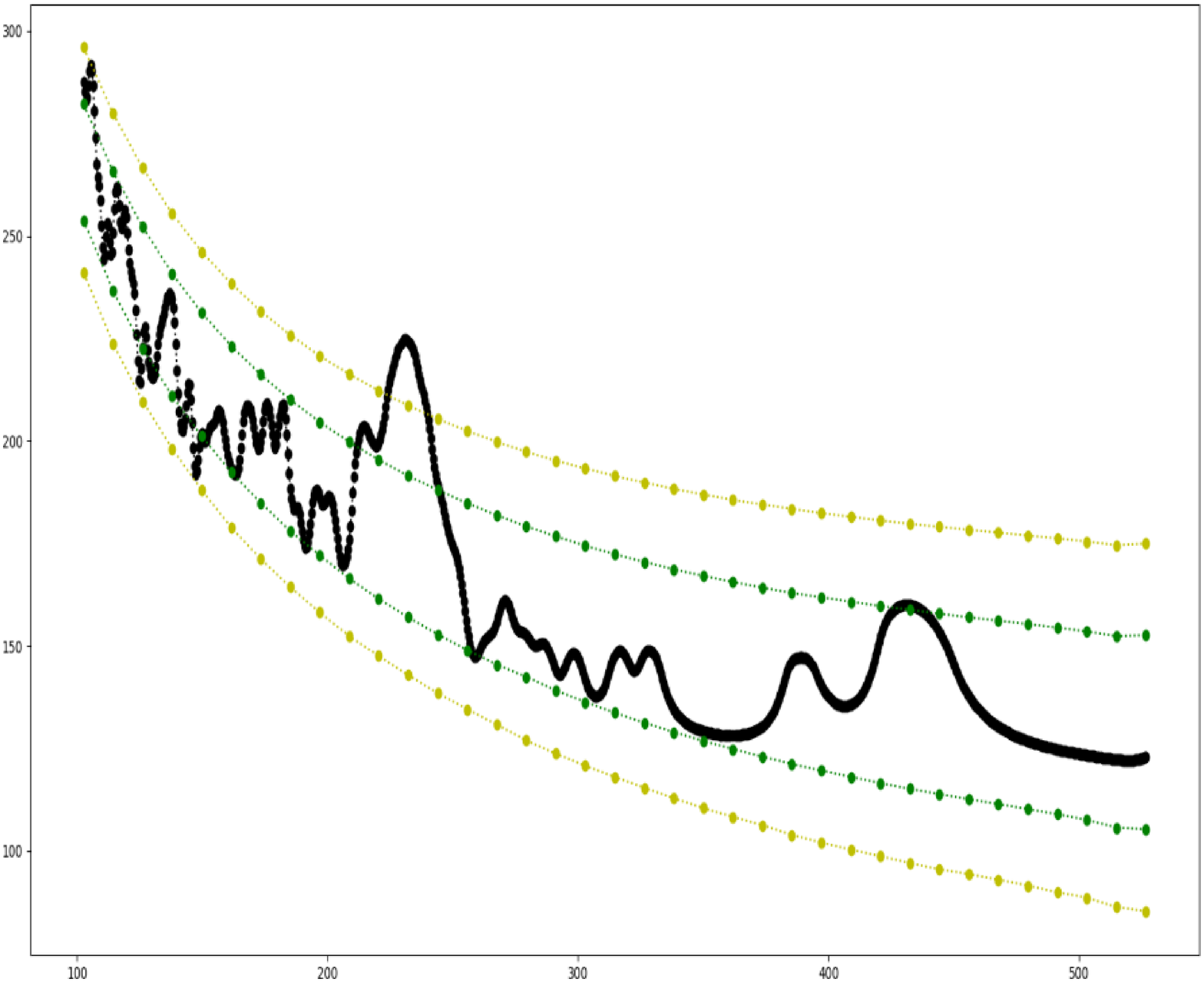}
\caption{}
\end{subfigure}
\end{figure*}

 After extracting the paths for the boundaries of the $1\sigma$ and
$2\sigma$ bands and the paths for the observed exclusion data points
(Figure~\ref{fig:exclusion}), the rest of the analysis
reduces down to comparison of $y$-coordinates of the data points with those
of the color boundaries in order to decide where each data
point lands: $1\sigma$/$2\sigma$ regions or beyond. The coordinate
information is extracted from the path properties.


There is no matching point on the sigma bands for every point on the experimental data line so only the points which have matching counterparts on the three curves are taken into consideration.

\subsection{Bitmap Analysis}

For the bitmap analysis, we convert the original versions of the
exclusion plots that have been collected from the article documents
into the PNG format. The PNG files have R,\,G,\,B,\,A values for each pixel
that we read using using Pillow~\cite{pillow2017}, a fork of the
lightweight Python Imaging Library (PIL).

This analysis depends on successfully identifying the observed limit curve and determining how many of its pixels are surrounded by green, yellow or white pixels. Given the myriad shapes of the curves and 1$\sigma$/2$\sigma$ color bands, some form of color and pattern recognition is needed. The most prominent characteristic of the observed exclusion curve is its black (or very dark) color. In order to make use of this feature, we start by removing other black (or dark) objects (such as letters written in black font or the little ticks placed on the plot axes) from the plots and then scan each column of pixels for remaining clusters of green, yellow and black pixels. These steps are detailed below:

\begin{itemize}
\item Find the black rectangular frame that is formed by the axes. Crop
  the image and leave only the parts falling inside of this frame.
\item Get rid of all the black objects (the numbers, letters, each dash of
  the dashed expected limit curves, etc.) except for the observed limit
  curve. For this, we first use the \texttt{findContours} function from
  the OpenCV library~\cite{opencv2018} to find all the connected regions
  (contours) in the image~\cite{suzuki1985}.  Any contours which could
  fit into a small rectangle (a rectangle with area smaller than 900 pixels)
  is removed from the plot (by coloring them blue, a neutral color that does
  not otherwise appear in the images).
\item Get rid of legend boxes (rectangular areas in yellow and green) so
  that the $1\sigma$ and $2\sigma$ bands are the only remaining yellow and
  green colored regions in the plots.  This is performed by a custom-developed
  function which identifies rectangles that run parallel to the frame
  borders and paints them blue.
\item As the final step, scan every column of pixels one by one. In each
  column find the group of black pixels (there should remain only one
  such cluster remaining per column), and examine the color of its
  neighboring pixels. If both neighbors of the black cluster in a
  given column is green (yellow), count that column as one data point
  within the $1\sigma$ ($2\sigma$) band.  In case one neighbor is
  yellow and the other is green, count that column as contributing half
  a datapoint in both bands.  And if one neighbor is yellow and the other
  is white, count that column as contributing half a datapoint only to
  the $2\sigma$ band.
\end{itemize}

On average about 430 columns are handled (and hence 430 data points are extracted from the limit curves) this way for each processed plot. For example for the plot seen in Figure~\ref{fig:exclusion}(a), 416 data points are obtained, with 32 located inside the white regions (outside of the $2\sigma$ bands); 10 located on the border of white and yellow regions (contributing 5 `weighted' points to the $2\sigma$ band); 55 located in one of the yellow regions (clearly inside the $2\sigma$ bands, but outside of the $1\sigma$ band); 37 located on the border of yellow and green regions (contributing 18.5 `weighted' points to the $1\sigma$ band); and 282 located in the green band (clearly inside the $1\sigma$ band). Hence the extracted fractions of points in the $1\sigma$ ($2\sigma$) region is 72.24\% (91.11\%).

\section{Results and Discussion}

Running over all our dataset, the vector graphics analysis extracts a
total of 7079 (5527) data points from the ATLAS (CMS) publications.
However before obtaining the cumulative results, we perform one final
step: We average out over images that are most likely to be highly
correlated. In the publications, the collaborations occasionally show
exclusion plots obtained from one particular set of data, but
interpreted for modified versions of the signal Monte Carlo
samples. (A good example is 1606.03833, which contains multiple copies of
Figure~\ref{fig:exclusion} prepared for various different values of
the decay width of resonance $X$.) In order not
to bias our meta-analysis, we weigh the data points extracted from
a set of such highly-similar figures in such a way that they
collectively get treated as if they are coming from one single
figure. After this treatment, the
effective count of data points is roughly halved.

\begin{table}[hbt!]
\centering
\topcaption{\label{tab:vecResults}Results from the vector graphics analysis.}
\begin{tabular}{|c|c|c|}
\hline*    
 & ATLAS & CMS \\\hline
Total count of data points &  7079  & 5527 \\ \hline
Effective count of data points after handling similar plots & 4157.33 & 2206.75 \\ \hline
Fraction of points in the 1\(\sigma\) band & 68.98\% \(\pm\) 0.72 & 74.04\% \(\pm\) 0.93 \\ \hline
Fraction of points in the 2\(\sigma\) band & 96.03\% \(\pm\) 0.30 & 95.23\% \(\pm\) 0.45 \\
\hline*    
\end{tabular}
\end{table}

The cumulative results of the vector analysis can be found in Table~\ref{tab:vecResults}. While the fraction of points within the $2\sigma$ region is consistent with 95\% for both experiments, the $1\sigma$ results are somewhat higher than 68\%. The bitmap analysis results are not yet ready for the CMS plots, due to the non-standard use of colors in those publications. The ATLAS results from the bitmap analysis (68.5\% and 94.1\%) are in good agreement with the vector graphics results.

Unfortunately, for these preliminary results, we are unable to provide
reliable uncertainty estimates yet: the values we report in the table are
simple binomial errors. We believe any systematic uncertainties due to
our ``re-digitization'' will be negligible in the vector graphics analysis,
as the collaborations produce the plots using ROOT with exact fidelity to
the actual limit values. We have performed a quick sanity check by
identifying the original values of the exclusion limits in the HEPData
repository~\cite{hepdata} for a couple of the publications and comparing
the results produced by our code with a manual counting performed on HEPData
points. We have observed excellent agreement. (We would like to use this
opportunity to invite the collaborations to be more generous with
HEPData. We could identify only 1 ATLAS and 7 CMS publications from
our list in the repository.)

In addition to the correct determination of the positions of the data points,
an essential component of the metaanalysis is the proper treatment
of any correlations between them. Beyond handling of the correlations
amongst similar plots in each paper, which we have described in the first
paragraph of this section (resulting in the decrease in the
effective count of points), we have considered two other sources of
possible correlations:

\begin{itemize}
\item \textit{Correlations over publications that use
  overlapping analysis selections and datasets.} As a very striking
  example, both of 1606.03977 and 1706.04786 are ATLAS searches
  for $W^\prime$ candidates in final states with a charged lepton
  and missing transverse momentum. The latter paper appears to describe the
  same kind of analysis as the former but run on a superset of the data.
  Our default treatment is to neglect all cross-paper correlations, since
  either the amount of data, or the generated Monte Carlo samples,
  or the analysis technique, or the list of specific final states
  vary significantly enough to keep the correlations weak. While
  we believe this is mostly justified, we have also implemented an alternative
  treatment: Using python, we extract ``refers-to'' lists for each of the
  analysed publications from arXiv.org and create network maps that display
  which of a given experiment's publications cites any other publication of
  that experiment in our list. Most publications are singletons
  in the map, but some produce small clusters. Then we repeat our
  metaanalysis either by averaging of results for members
  of each cluster, or by removing publications from our list until the
  maps consist of singleton papers only. The second method removes 4 CMS
  and 8 ATLAS publications, but we are still left with over 5000 data
  points for each experiment. The final results obtained either way are
  compatible with the results reported in Table~\ref{tab:vecResults}.
\item \textit{Correlations of data points in each plot.}
As the LHC collaborations use one set of Monte
Carlo background events for all the different values of a given signal's
paramaters (like a new particle's mass) and since those parameters have
resolutions possibly larger than inter-point separations in the plots, the
closeby data points on each plot are likely to be somewhat correlated.
We intend to carefully model such correlations in a future
publication. Yet, as a quick self-test, we have repeated the analysis by
randomly picking no more than 3 data points per plot, with each point
located far from each other to keep the correlations negligible. The
results are again compatible with those reported in
Table~\ref{tab:vecResults}, but with much larger uncertainties due to
limited statistics. Such an analysis will become more interesting as we
include more papers and plots in the future.
\end{itemize}

As a conclusion, despite the minor issue that our metaanalysis has
identified, the overall results from both ATLAS and CMS experiments are
quite encouraging. In a world where the scientific endevour has experienced
recent reproducibility issues, we are happy to see that the results from
the field of experimental high energy physics appear to be sound. We intend
to improve the analysis and make it semi-automatic to keep monitoring
new publications.

\section*{Acknowledgements}
We would like to thank the anonymous referree for her/his careful reading of
the manuscript and providing ideas for substantial improvements of the
present text and for the future of the analysis.

\end{document}